\documentstyle[aps,preprint,graphicx]{revtex}

\begin{document}
\draft

\title{Electronic structure of wurtzite and zinc-blende AlN}

\author{P. Jonnard \and N. Capron}
\address{Laboratoire de Chimie Physique - Mati\`ere et Rayonnement,
Universit\'e Pierre et Marie Curie, UMR-CNRS 7614, 11 Rue Pierre et 
Marie Curie, F-75231 Paris Cedex 05, FRANCE.}

\author{F. Semond \and J. Massies}
\address{CRHEA/CNRS, rue B. Gregory, Sophia Antipolis, F-06560 
Valbonne, FRANCE.}

\author{E. Martinez-Guerrero \and H. Mariette}
\address{CEA-CNRS research group "Nanophysique et Semiconducteurs",
Laboratoire de Spectrom\'etrie Physique, Universit\'e J. Fourier,
CEA/DRFMC/SP2M, 17 avenue des Martyrs, F-38054 Grenoble, FRANCE.}

\date{\today}
\maketitle

\begin{abstract}
The electronic structure of AlN in wurtzite and zinc-blende phases is 
studied experimentally and theoretically. By using x-ray emission 
spectroscopy, the Al~$3p$, Al~$3s$ and N~$2p$ spectral densities are 
obtained. The corresponding local and partial theoretical densities of 
states (DOS), as well as the total DOS and the band structure, are 
calculated by using the full potential linearized 
augmented plane wave method, within the framework of the density 
functional theory. There is a relatively good agreement between the experimental 
spectra and the theoretical DOS, showing a large hybridization of the valence 
states all along the valence band. The discrepancies between the experimental 
and theoretical DOS, appearing towards the high binding energies, are 
ascribed to an underestimation of the valence band width in the 
calculations. Differences between the wurtzite 
and zinc-blende phases are small and reflect the slight 
variations between the atomic arrangements of both phases.
\end{abstract}


\narrowtext

\section{Introduction}
The III-nitrides are nowadays widely used by the semiconductor 
industry. With respect to "classical" III-V semiconductors, the 
nitrides have larger band gaps and stronger bonds. This makes them 
attractive for optoelectronic devices that need short 
wavelength of the visible range (violet and blue regions), for 
devices that work at high temperatures or with high powers. Among them, 
AlN has the largest band gap (6.2~eV) and thus, is the best 
material for constructing devices for the violet region. 
Moreover, with GaN, it can form Ga$_{x}$Al$_{1-x}$N
ternary alloys, with $0 \leq x \leq 1$. This enables the 
tuning of the emission wavelength of the optoelectronic devices by 
choosing a suitable composition of the alloy.

AlN can be grown in wurtzite (WZ) or zinc-blende (ZB) phase. The 
stable one is the WZ phase, while the ZB phase is metastable. The 
optical gap of the WZ phase is direct and that of the ZB phase is 
indirect. This can be useful in constructing different kinds of quantum wells 
or superlattices. However, the growing of AlN is still a difficult 
task~\cite{jai}, particularly for the ZB phase. Due to the lack of good 
quality samples, only little work was done to study the 
physical properties of both phases. In the present work, we 
experimentally study the electronic structure of well characterized samples 
of both AlN phases, and compare the results with our 
theoretical calculations of the valence density of states (DOS).

The experimental electronic structure is obtained by 
x-ray emission spectroscopy (XES) induced by electrons. XES has 
been widely used to study the electronic structure of 
III-V semiconductors~\cite{ver}. The DOS are 
calculated by using the full potential linearized 
augmented plane wave method, within the framework of the density 
functional theory (DFT)~\cite{hoh}. The experimental and theoretical 
determination of the various partial and local DOS is important for 
the study of the hybridization of the states inside the valence band.

\section{Experimental details}
\subsection{Sample preparation}
\subsubsection{Wurtzite AlN}
A 300 nm thick wurtzite AlN film is grown by molecular beam epitaxy 
(MBE) in a RIBER Compact 21 system, using ammonia 
as the N precursor, and standard solid sources for the group-III elements. 
The substrates is a 2'' diameter Si~(111) 
wafer with no ex-situ surface preparation. The in-situ silicon wafers cleaning 
involves a prolonged heating (about 10~hours) at 600$^{\circ}$C with subsequent removal
of the oxide layer by rapid thermal annealing up to about 1000$^{\circ}$C. As followed 
by reflection high energy electron diffraction (RHEED), the appearance 
of the ($7 \times 7)$ surface reconstruction indicates that the surface is clean and well 
ordered. The AlN layer is grown at about 920$^{\circ}$C. The set of growth parameters 
(mainly growth temperature and growth rate) is properly adjusted in order to rapidly 
obtain a layer by layer growth. This is achieved by using a very low growth 
rate (0.1~mm/h),
and a nucleation temperature of 650$^{\circ}$C followed by a rapid ramping to the growth 
temperature (920$^{\circ}$C). The AlN RHEED pattern evolution during growth indicates 
that the growth first follows a three-dimensional mode, and progressively becomes 
two-dimensional (2D) after some tenths of nanometers. By decreasing the temperature after 
the growth, a $2 \times 2$ surface reconstruction characteristic of the Al polarity is observed. 

The structural quality of AlN films is investigated by transmission electron microscopy 
(TEM), x-ray diffraction (XRD) and atomic force microscopy (AFM). Measurements of the 
\textit{rms} roughness by AFM give values of 0.4~nm. Cross-sectional TEM images show that the 
interface between AlN and the silicon substrate is sharp, and no amorphous interfacial layer 
can be observed. However a very high threading and misfit dislocations density is present 
due to the large lattice mismatch between AlN and Si. The dislocation density in the AlN 
layer is estimated from plan-view TEM images, and is about $5 \times 10^{11}$~cm$^{-2}$. 
During the cooling after the growth, an extensive stress appears due to the large 
thermal mismatch between Si and the AlN layer. This is confirmed by XRD 
measurements, which show an out-of-plane $\epsilon_{zz}$ deformation of Ð2.2\%.

\subsubsection{Zinc-blende AlN}
The sample is grown by plasma-assisted MBE. The metal 
fluxes are provided by conventional effusion cells, while active nitrogen 
results from radio frequency 
dissociation of N$_{2}$ using a plasma cell. The used pseudo-substrate consists of a 3~$\mu$m
thick 3C-SiC layer grown by chemical vapor deposition on a Si(100) 
substrate~\cite{cha}. To follow the evolution of the surface morphology, 
the growth is monitored in situ by RHEED. 
A smooth 2D-layer of zinc-blende AlN, with a thickness of about 300~nm, is obtained for 
a growth temperature of 720$^{\circ}$C under stoichiometric conditions. 
The roughness profile of AlN, deduced from (1~mm $\times$ 1~mm) topographic AFM images, 
shows superficial steps with a maximum height of 0.8~nm. The structural quality of 
the AlN buffer layer is checked  by high resolution x-ray diffraction 
study~\cite{mart}, and the purity of the cubic crystallographic phase is assessed 
by Raman spectroscopy measurements~\cite{bus}.

\subsection{X-ray emission spectroscopy}
\subsubsection{Studied emissions}
In XES, one observes the radiative transitions towards 
a hole present in a core shell. The energy distribution of the emitted 
photons (called spectral density) is proportional to the convolution of the 
DOS of the initial and final states, weighted by 
the matrix element of the transition probability, and by the cube of the photon 
energy~\cite{bon1}. If the transition involves the valence 
band, the observed spectral density describes the valence DOS, 
broadened by a Lorentzian that represents the lifetime broadening of the 
initial state (i.e. a hole in a core shell) and 
a bell shape function representing the instrumental function. 
This is true only if the radiative transition probability does not 
vary strongly across the valence band, which is generally a good 
assumption. Because of the local character of the core hole wavefunction 
describing the initial state and of the dipole selection rules 
($\Delta l = \pm 1$), the observed DOS are local, \textit{i.e.}, in a 
compound the DOS is decomposed for each element, and 
partial, \textit{i.e.}, each $l$-like symmetry of the DOS is obtained separately.
 
To study the occupied valence states of AlN, the following emission 
bands are analyzed~:
\begin{itemize}
\item[-] the Al~K$\beta$ emission ($3p \to 1s$ transition), describing the Al~$3p$ states;
\item[-] the Al~L$_{2,3}$ emission ($3sd \to 2p$ transition), describing the Al~$3s$ and $3d$ states;
\item[-] the N~K$\alpha$ emission ($2p \to 1s$ transition), describing the N~$2p$ states.
\end{itemize}

When compared with the theoretical DOS, the experimental spectral 
densities are 
set on a binding energy scale relative to the top of the valence 
band, E$_{V}$. This energy cannot be deduced from the combination of 
the XES experiments and the binding energies of the various core levels 
involved in the emissions. Indeed, the binding 
energies are determined by photoemission with respect to the Fermi 
level, E$_{F}$, whose position with respect to E$_{V}$ is not 
known, and some states present in the AlN optical gap can induce a 
large shift of E$_{F}$. The binding energy scale for the spectral 
densities is obtained by adjusting the main feature of the experimental 
and theoretical DOS, leading 
to an uncertainty of $\pm$0.3~eV in the position of E$_{V}$. For the 
ultra-soft x-ray emissions, the variation of the photon energy must be
taken into account. Then, the Al~$3sd$ spectral density is divided by 
the cube of the photon energy. For the 
other emissions this correction is small enough to be omitted. 
 
\subsubsection{Experiment}
The Al~K$\beta$ emission is obtained using a curved crystal 
Johann-type x-ray spectrometer~\cite{bon2}, with a ${(10\overline{1}0)}$ 
quartz crystal curved with a radius of 500~mm. The spectral 
resolution, $E/\Delta E$, is about 3000. The Al~L$_{2,3}$ and 
N~K$\alpha$ emissions are obtained using a 2~m grazing incidence (1.5$^\circ$) 
x-ray spectrometer~\cite{jls}, with a 600~grooves/mm grating. 
The N~K$\alpha$ emission is obtained in the third order of diffraction. 
The spectral resolution is about 1000 for the N~K${\alpha}$ photon energy 
range, and about 750 for the Al~L$_{2,3}$ photon energy range.

The Al~L$_{2,3}$ and N~K$\alpha$ emissions are induced by a 3~keV electron 
beam, and the Al~K$\beta$ emission by a 4~keV electron beam. Under these 
conditions the analyzed thickness is around hundred nanometers. 
For such a thickness, the samples can be considered as bulk. Since the electron 
current densities employed are low (about 1~mA/cm$^{2}$), no damage can be 
induced inside the sample by the electron beam. Indeed, we 
verify that no evolution in time of the spectral shape 
and intensity occurs during data acquisition.

\section{Computational details}
The calculations are performed using the full-potential 
linearized-augmented-plane-wave method, within the framework of the DFT, 
as implemented in the WIEN2k package~\cite{w2k}. This method has 
proved successful in the determination of the electronic 
structure of AlN~\cite{lit}. Generalized gradient approximation (GGA) 
corrections to the local exchange-correlation functional 
are used as proposed by Perdew-Burke-Ernzerhof~\cite{pbe}. 
The GGA has been successfully used to study different phases of ionic 
compounds~\cite{ham,sta}.

The atomic structure is fixed to the experimental one for both 
phases. No geometry optimization is performed. The muffin-tin radii 
of the N and Al spheres are chosen as 1.677 and 1.877 Bohr, 
respectively. The $k$-integration over the Brillouin zone is performed 
on a $19 \times 19 \times 10$ mesh for the WZ phase, and on a 
$16 \times 16 \times 16$ mesh for the ZB phase, leading to 400 and 
446 irreducible points in the Brillouin zone, respectively.

\begin{figure}
\includegraphics[scale=0.75]{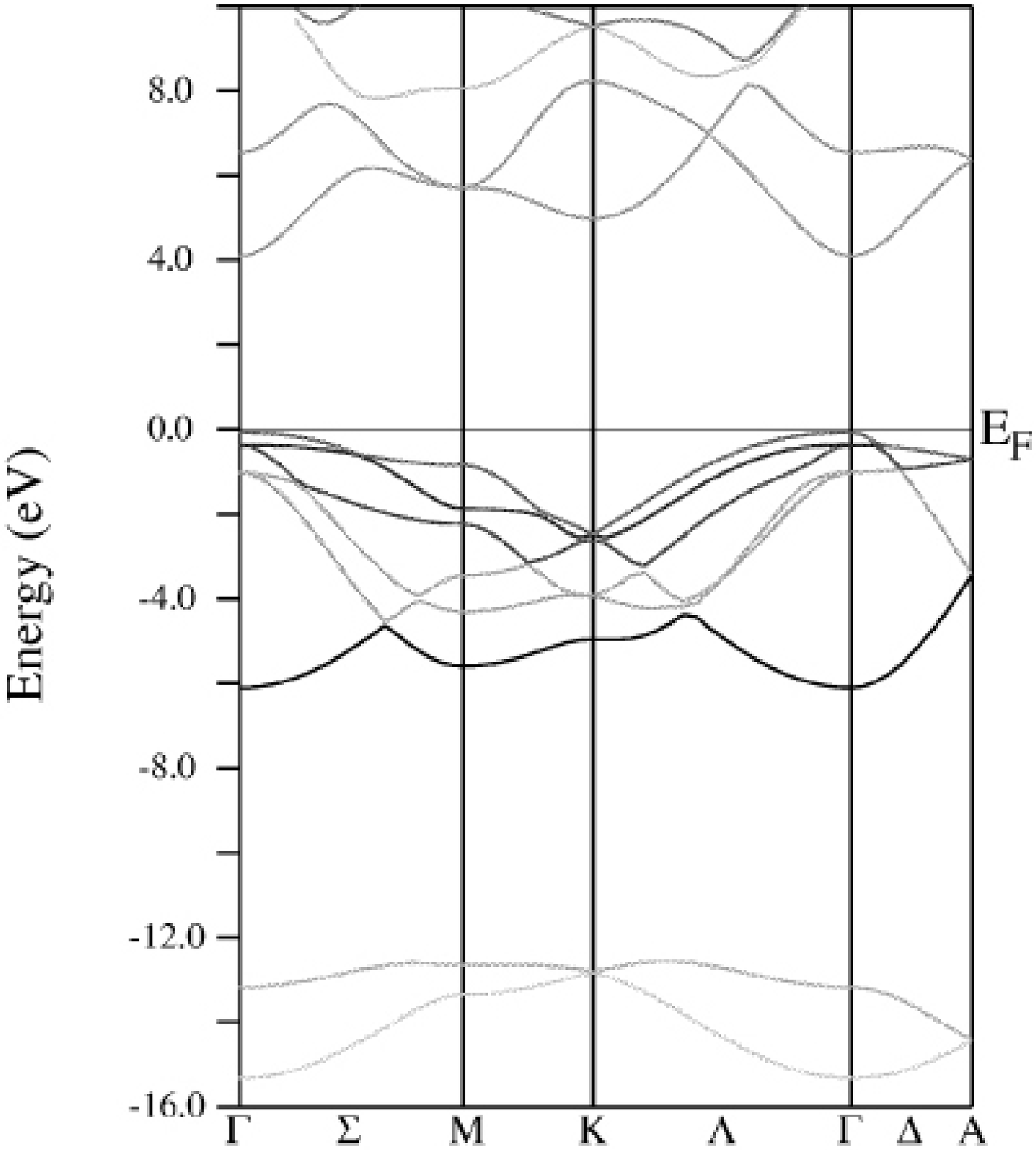}
\caption{Band structure of AlN WZ.}
\label{AlNwBS}
\end{figure}

\begin{figure}
\includegraphics[scale=0.75]{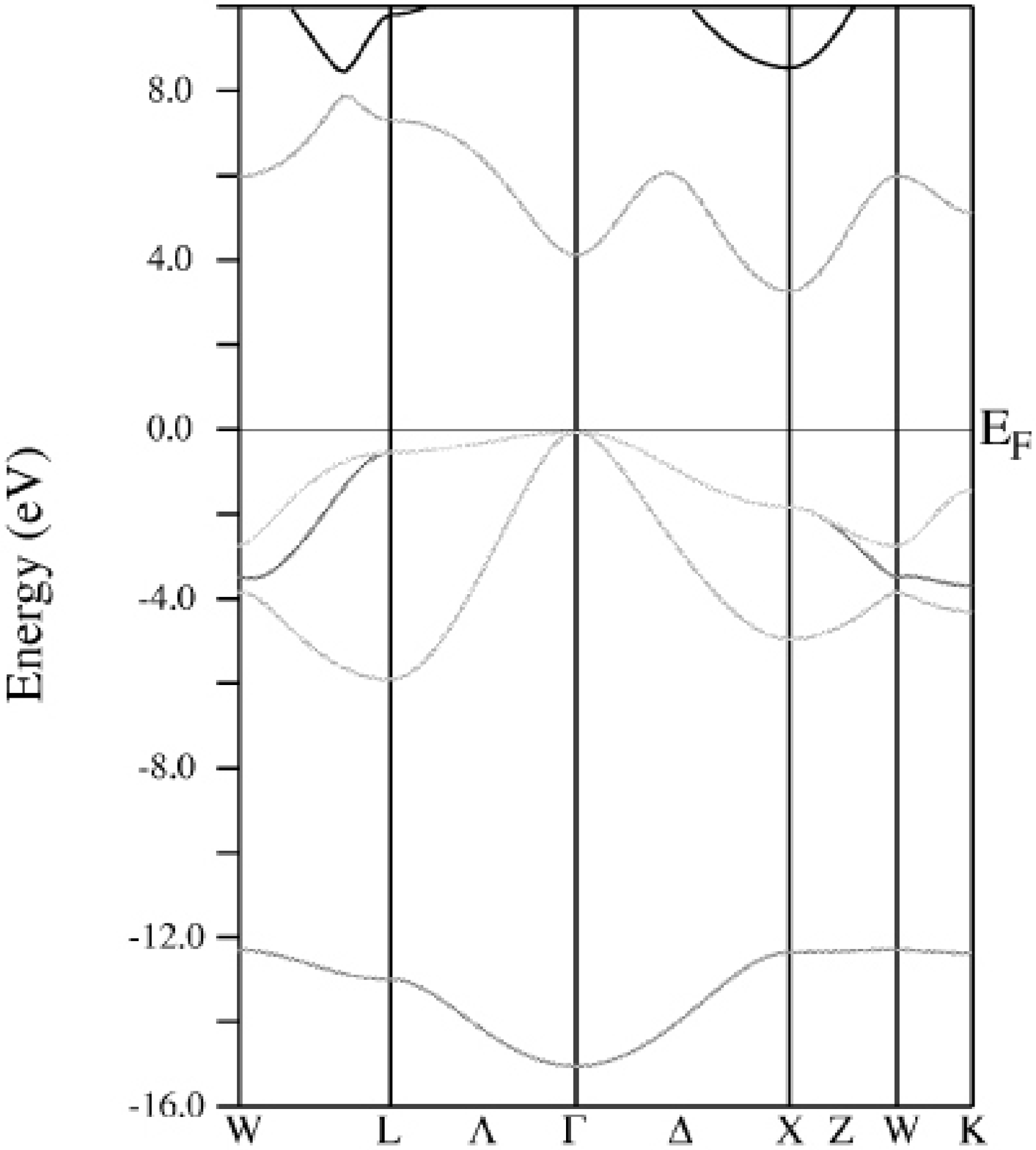}
\caption{Band structure of AlN ZB.}
\label{AlNcBS}
\end{figure}

The calculated band structures for the WZ and ZB phases are presented 
in the Figures~\ref{AlNwBS} and ~\ref{AlNcBS}, respectively. It can be 
seen that the gap of WZ AlN is direct and that of ZB AlN is indirect, 
between the $\Gamma$ and $X$ points. This description is in agreement 
with calculations performed on both WZ and ZB phases, using the DFT 
and the local density approximation~\cite{lit,pug,mer}, or using an empirical 
pseudopotential approach~\cite{fri}.

\begin{figure}
\includegraphics[scale=0.75]{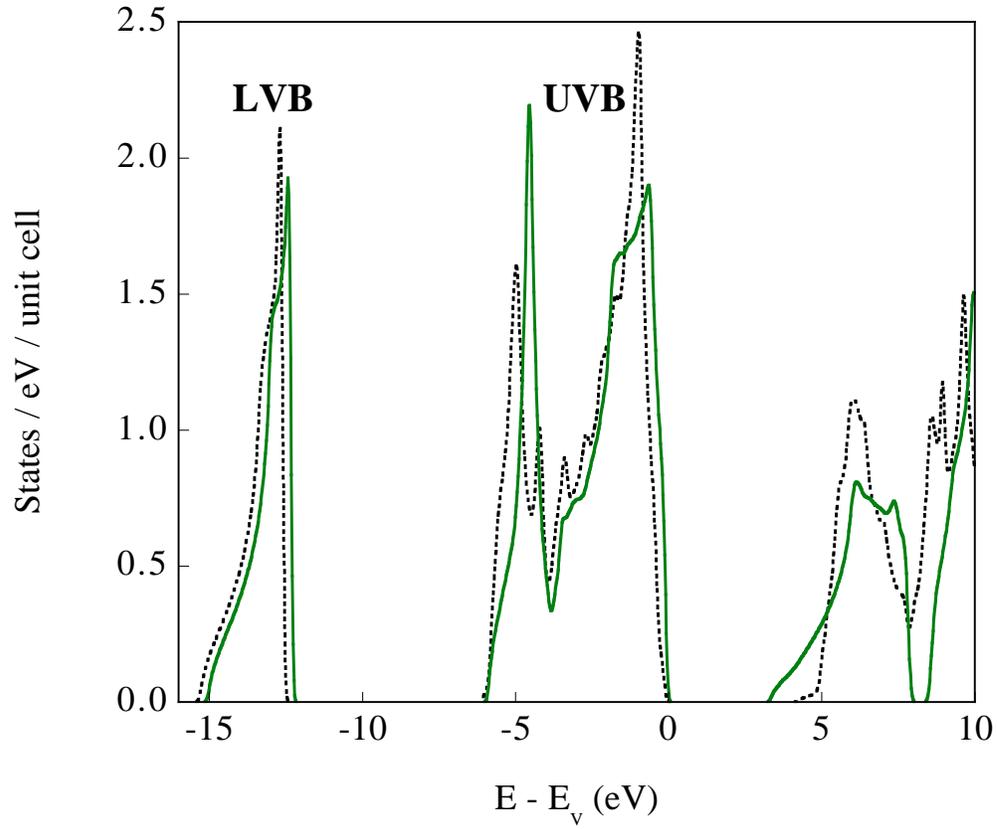}
\caption{Total density of states of AlN WZ (dotted line) and ZB 
(solid line).}
\label{DOStot}
\end{figure}

In Figure~\ref{DOStot} the total DOS for WZ and ZB 
AlN are presented on a binding energy scale relative to $E_{V}$. 
The calculated DOS are in agreement with previous calculations performed for 
these phases~\cite{klk,lit}. The occupied valence states are 
distributed in two parts : in the lower valence 
band (LVB) towards the high binding energies, and in the upper valence 
band (UVB) towards the low binding energies. The calculated optical band 
gaps are 4.4~eV and 3.1~eV for WZ and ZB AlN, respectively. As expected 
from DFT calculations, these values are somewhat lower than the 
experimental values of 6.2 and 5.4~eV~\cite{vur} for WZ and ZB AlN, respectively.

When compared to the XES experiments, the local and partial DOS are
\begin{itemize}
\item[-] weighted by the matrix element of the corresponding transition,
\item[-] broadened by a Voigt function to take into account the instrumental 
function,
\item[-] broadened by a Lorentzian function to take into account the lifetime 
of the core hole,
\item[-] broadened by a Lorentzian of variable width to take into account the 
variable lifetime of the hole in the valence band. The valence hole 
width, $W$, varies from 0~eV at $E_{V}$, 
to 1~eV at the bottom of the valence band, $E_{B}$, following~:
$$W=\left( \frac{E-E_{V}}{E_{B}-E_{V}} \right) ^{2}$$
where $E$ is the position of the hole within the valence band.
\end{itemize}
The DOS, which have been treated by this procedure, are called 
calculated spectra hereafter.

\section{Experimental results and calculations}
\subsection{Al $3p$ spectral densities}
\begin{figure}
\includegraphics[scale=0.75]{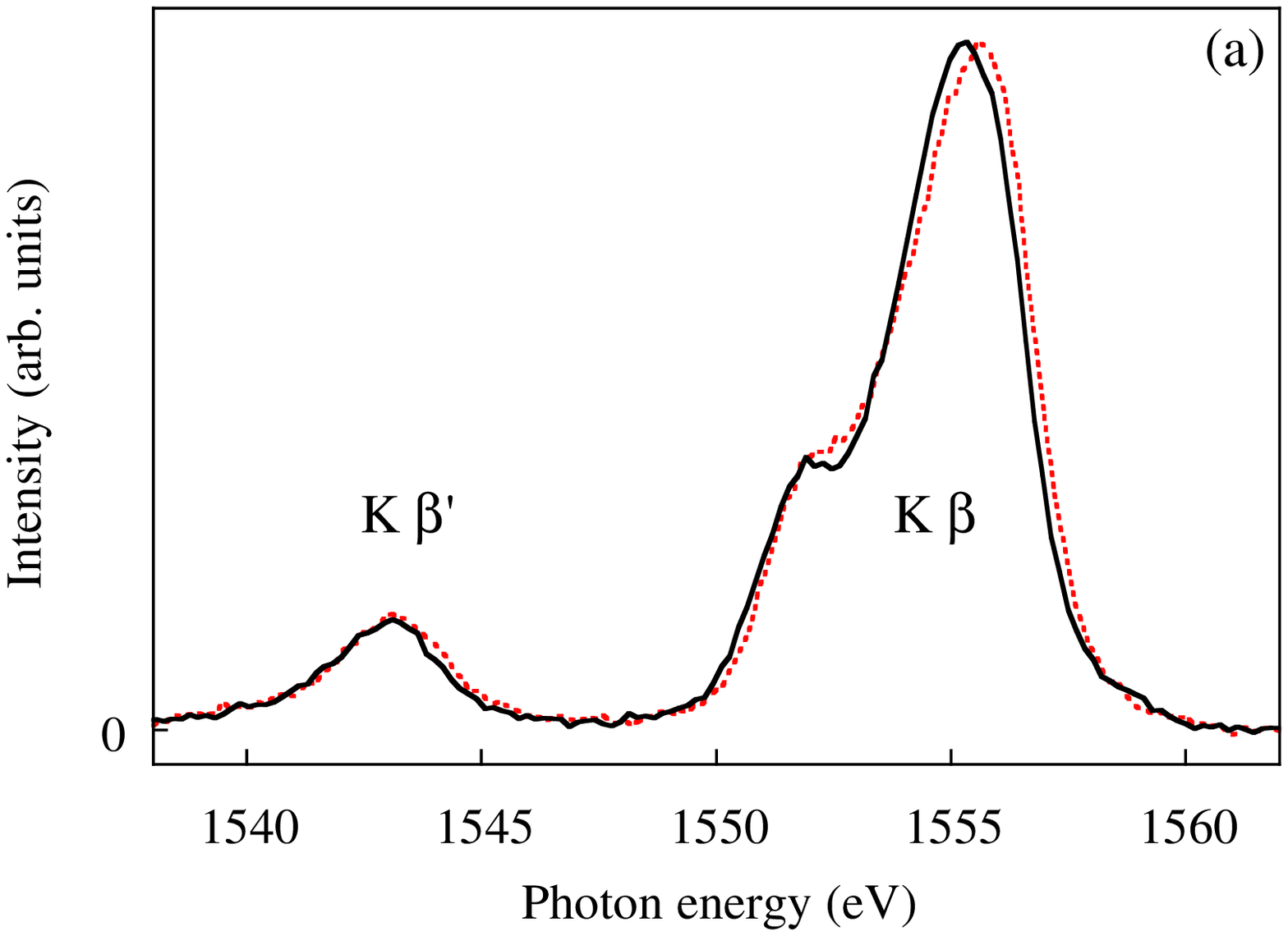}
\includegraphics[scale=0.75]{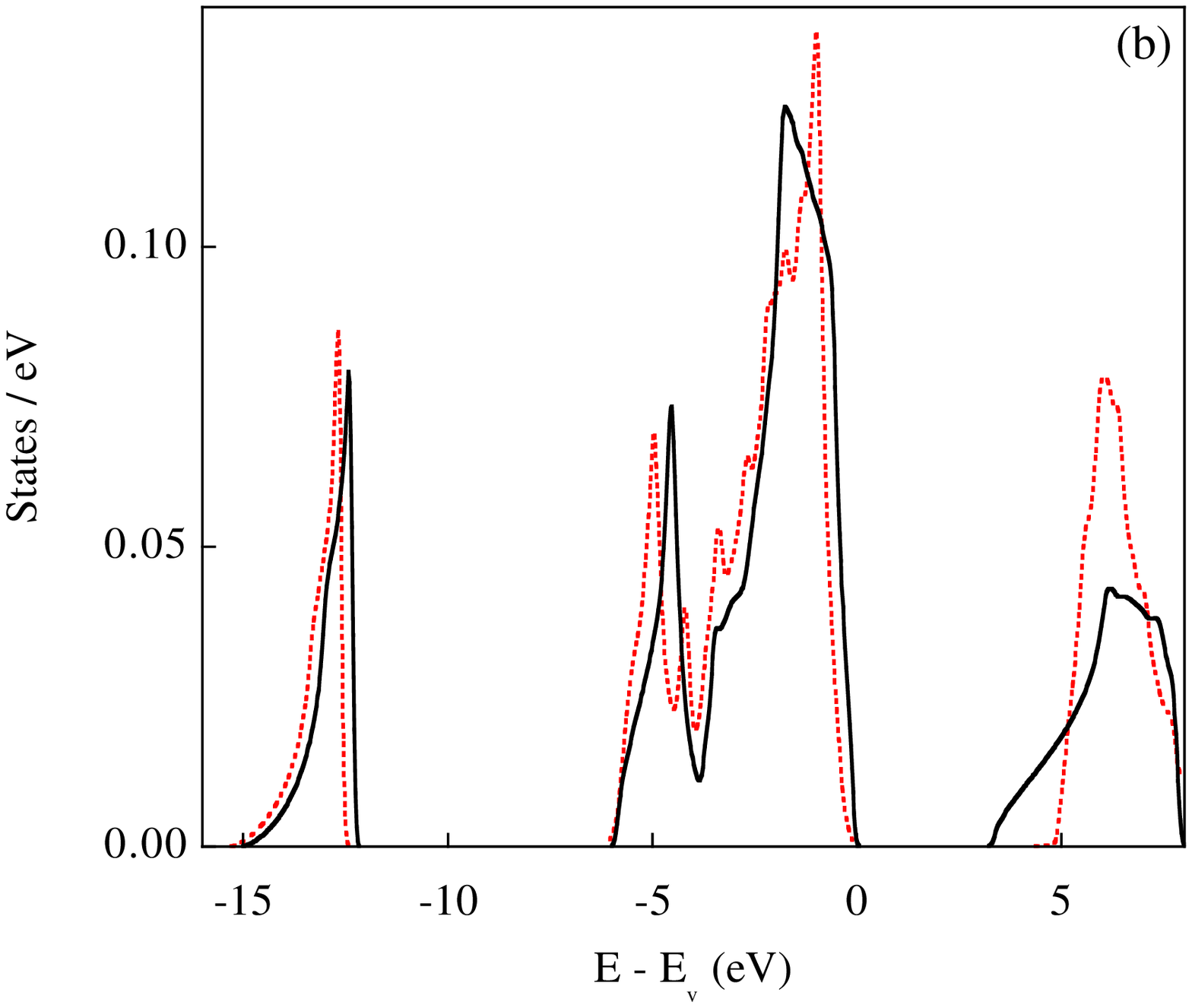}
\clearpage
\includegraphics[scale=0.75]{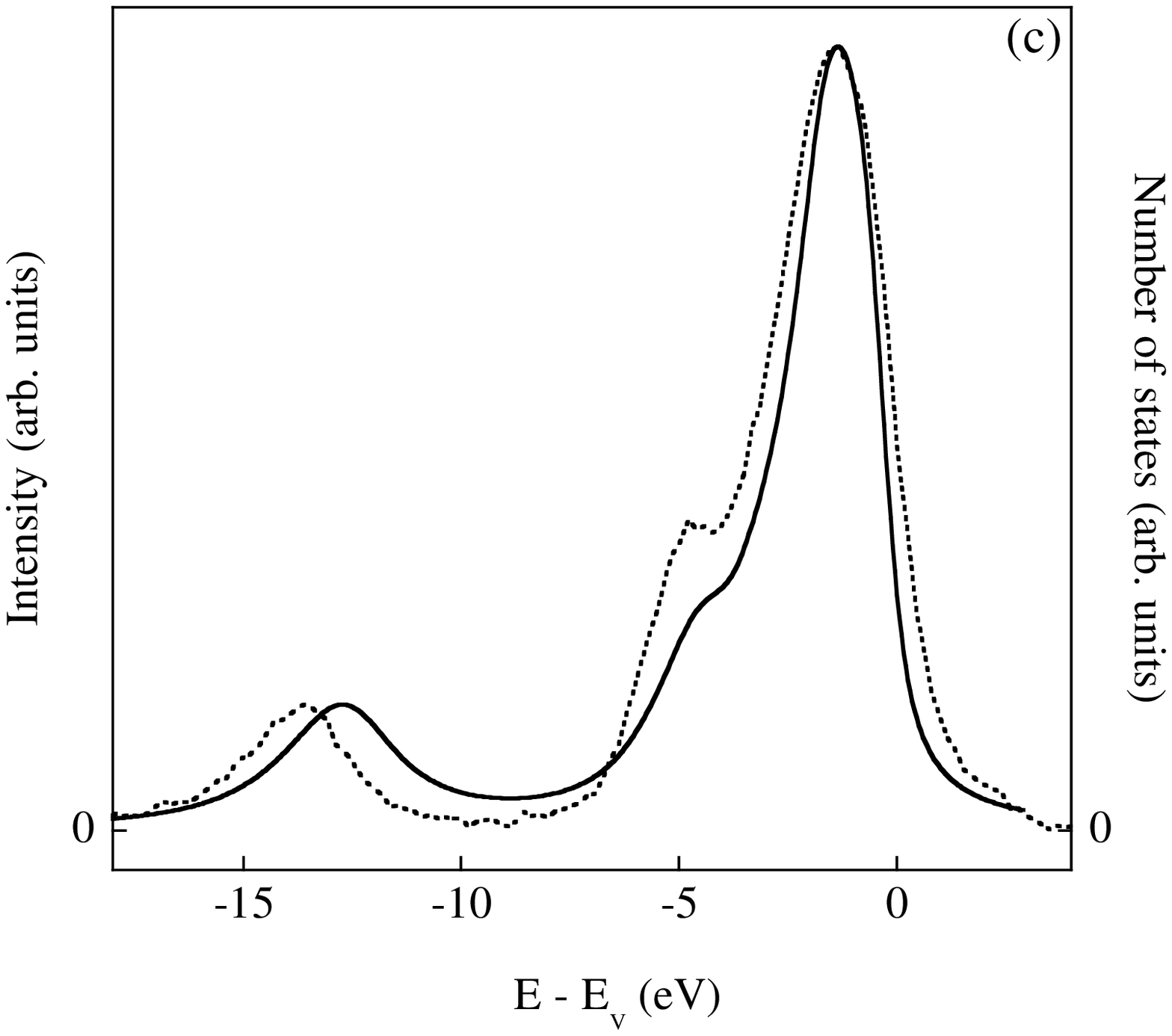}
\caption{(a) Al $3p$ spectral distribution of AlN WZ (dotted line) and 
ZB (solid line) obtained with 4~keV incident electrons; (b) Theoretical 
Al~$3p$ DOS of AlN WZ (dotted line) and ZB (solid line); (c) For AlN 
ZB, comparison of experimental (dotted line) and calculated (solid line) 
Al~$3p$ spectra.}
\label{Alk}
\end{figure}

The Al~$3p$ spectral densities of WZ and ZB AlN are presented 
in Fig.~\ref{Alk}a, on a photon energy scale. They are normalized with 
respect to their maximum.
They present a main peak (labeled K$\beta$) at about 
1555~eV with a shoulder around 1552~eV. A secondary peak (labeled 
K$\beta '$) is observed at about 1548~eV. The shape and relative 
intensity of the WZ phase is almost in perfect agreement with a 
previous experiment~\cite{wie}. Another experiment~\cite{rei} obtained 
with a poorer spectral resolution is also in agreement with the 
presented spectra. Two other experiments~\cite{fom,smi}
can be found in the literature, but they strongly disagree with the 
presented spectrum. Since they show supplementary structures, 
which cannot be explained by DOS calculations (see below), these 
spectra were probably obtained from poor quality samples.

The differences observed between the spectra of the two AlN phases are small,
Fig.~\ref{Alk}a. With respect to the WZ phase, the maximum and the top of the 
valence band of the ZB phase are 
shifted towards the low photon energies, by 0.3~eV and 0.2~eV, 
respectively. The shoulder is slightly less intense for the ZB 
phase. A small difference is also noted on the K${\beta}'$
structure, which is narrower towards the high photon energies for 
the ZB phase.

The theoretical Al~$3p$ DOS for AlN~WZ and ZB are presented in 
Fig.~\ref{Alk}b. The differences between the two phases are rather 
small in the occupied states. 
The calculations are in agreement with previous local and partial DOS 
calculated for AlN WZ~ and ZB~\cite{klk,lit}.

The comparison of the Al~$3p$ spectral density with the corresponding calculated spectrum
 is show in Fig.~\ref{Alk}c for AlN~ZB. It can be seen that the K$\beta$ 
and K$\beta '$ ranges correspond to the UVB and the LVB, respectively. 
There is a satisfactory agreement between the experimental and theoretical 
DOS. In particular, the energy difference between the main peak and its 
shoulder is well reproduced. However, their relative intensity is different
in the experiment and the theory. Another discrepancy
can be noted for the K$\beta '$ structure~: in the theoretical 
DOS, this structure is shifted by 1~eV towards 
the lower binding energies, with respect to the experimental structure. 
Comparison between the experimental and theoretical DOS of AlN~WZ 
leads to similar remarks.

\subsection{Al $3sd$ spectral densities}
\begin{figure}
\includegraphics[scale=0.75]{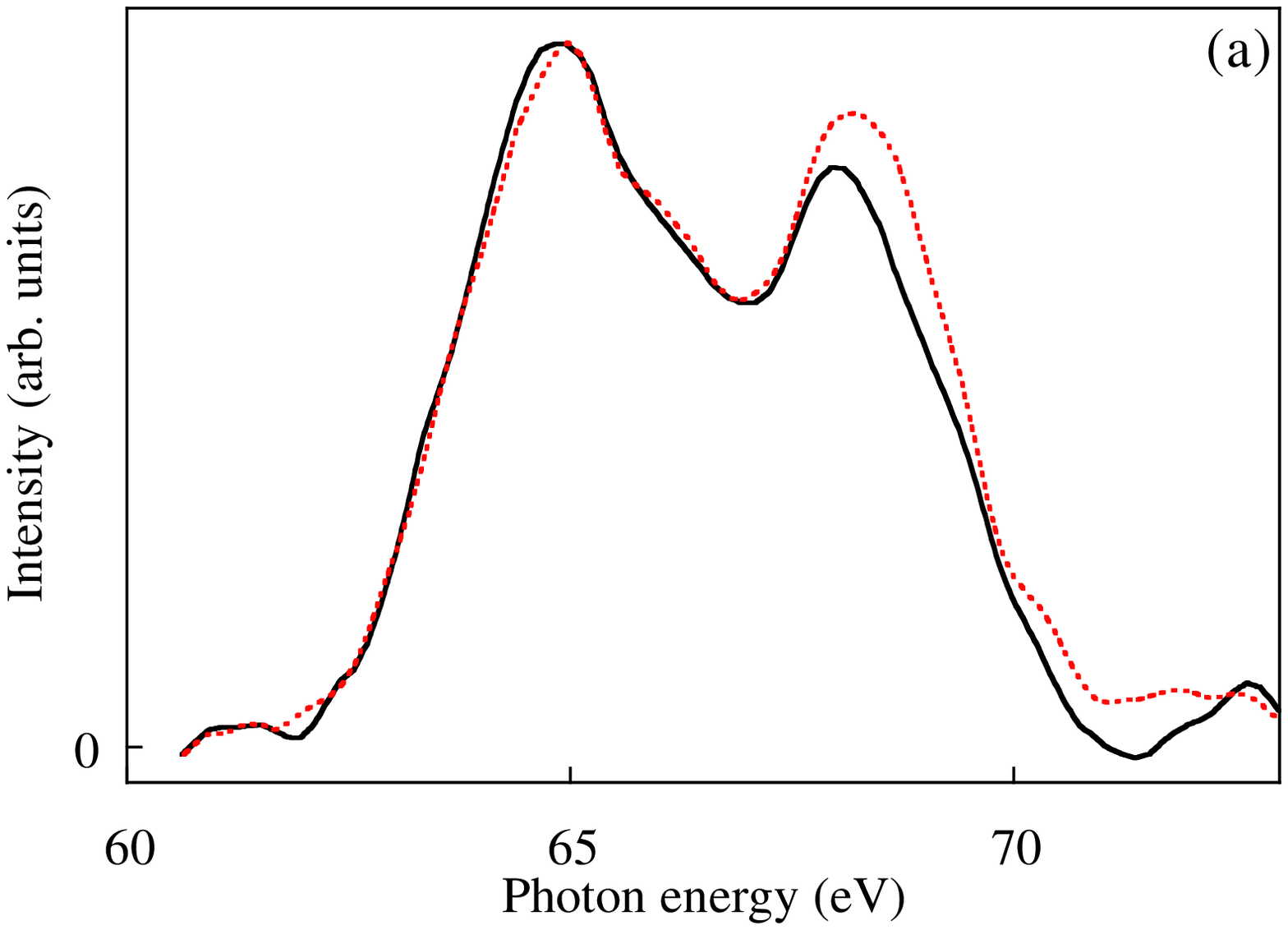}
\includegraphics[scale=0.75]{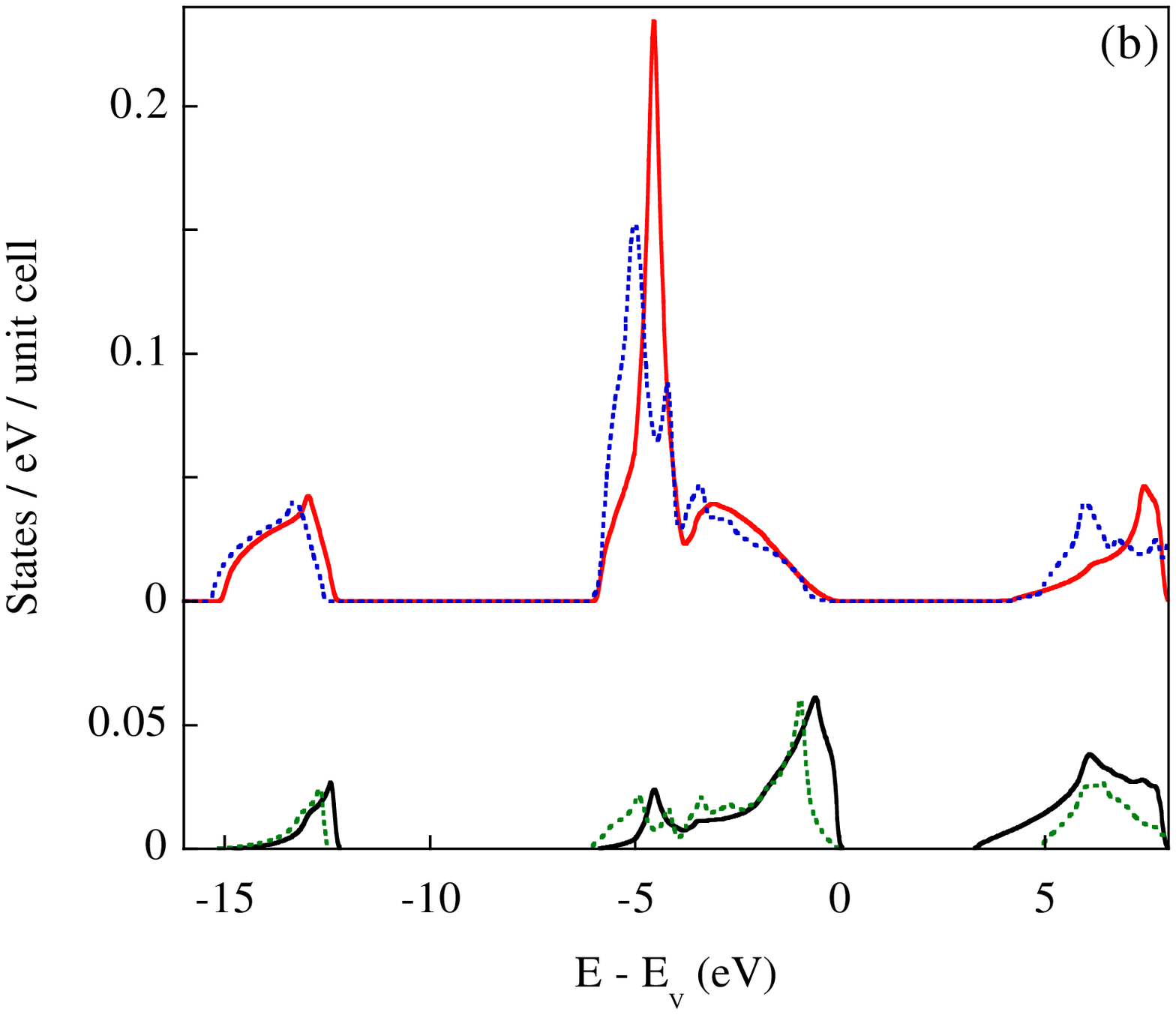}
\clearpage
\includegraphics[scale=0.75]{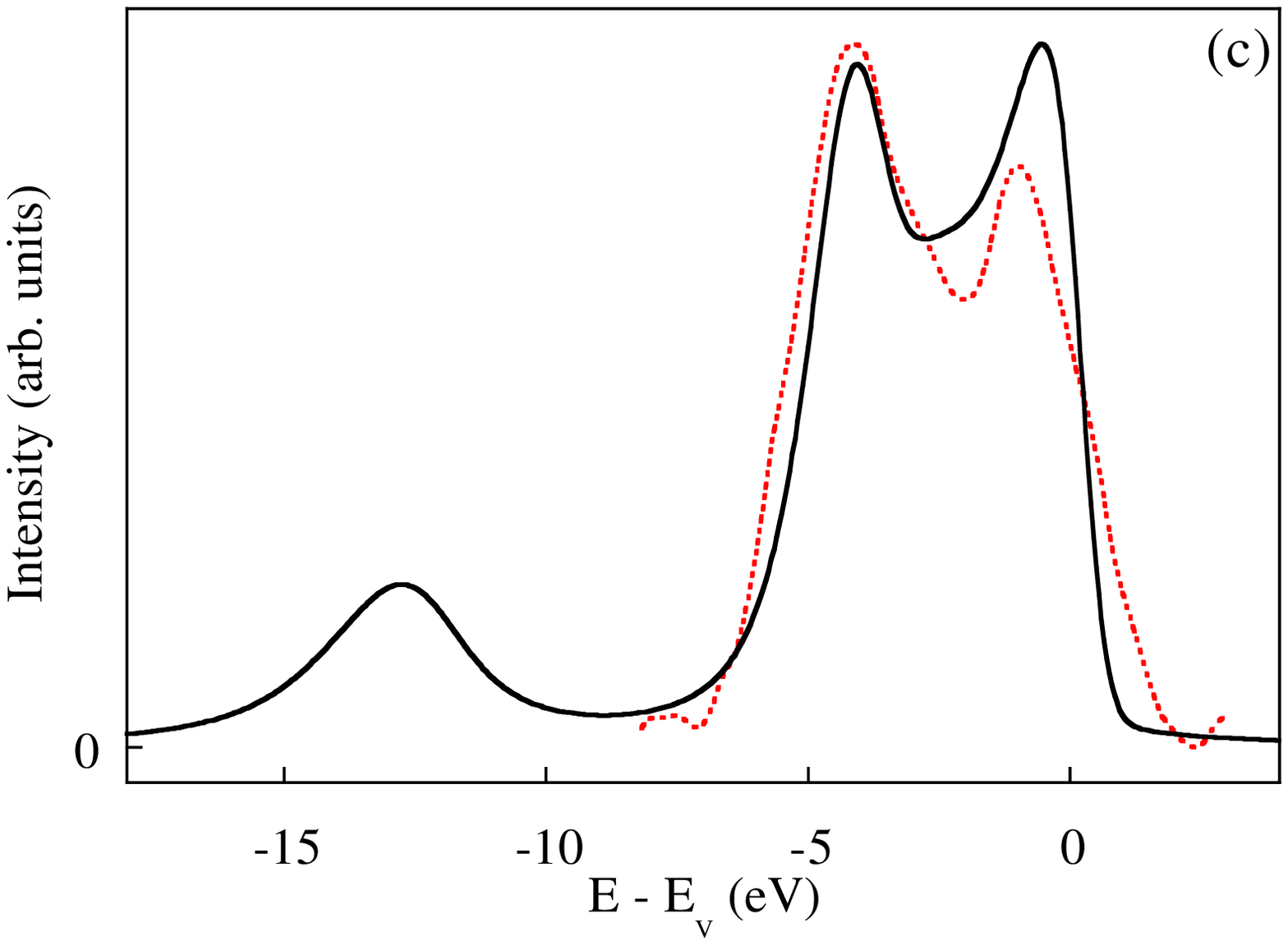}
\caption{(a) Al $3sd$ spectral distribution of AlN WZ (dotted line) and ZB 
(solid line) obtained with 3~keV incident electrons; (b) Theoretical 
Al~$3s$ (top) and Al $3d$ (bottom) DOS of AlN WZ (dotted line) and ZB 
(solid line); (c) For AlN ZB, comparison of experimental (dotted line) 
and calculated (solid line) Al~$3sd$ spectra.}
\label{AlL}
\end{figure}

The Al~$3sd$ spectral densities of WZ and ZB AlN are presented 
in Fig.~\ref{AlL}a, on a photon energy scale. 
They are normalized with respect to their maximum. 
They present a doubly structured peak, with maxima at 65 and 68~eV. The 
spectrum of the WZ phase is in very good agreement with that 
presented in Ref.~\onlinecite{wie}. Another spectrum of WZ AlN 
can be found in the literature~\cite{fom}, but with a poorer 
resolution. With respect to the WZ phase, the peak towards 
the high photon energies is of lower intensity for the ZB phase.

The theoretical Al~$3s$ and $3d$ DOS AlN~WZ and ZB are presented on a binding 
energy scale relative to the top of the valence band in 
Fig.~\ref{AlL}b. The calculations are in agreement with previous local 
and partial DOS calculated for AlN WZ~ and ZB~\cite{klk,lit}. 
The differences between both phases are small. It is deduced that the maximum 
at 65~eV on the spectral DOS corresponds essentially to Al~$3s$ states, 
whereas Al~$3s$ and $3d$ states contribute to the maximum at 68~eV.

For AlN ZB, the Al~$3sd$ spectral density is compared to the calculated 
spectrum in Fig.~\ref{AlL}c. It is observed that the positions of 
the two peaks agree in the experiment and the calculation, but the relative 
intensities of these peaks are not well reproduced by the 
calculation. However, it should be emphasized that the evolution of 
these relative intensities between the WZ and ZB phases, is correctly 
described by the calculation. 

\subsection{N $2p$ spectral densities}
\begin{figure}
\includegraphics[scale=0.75]{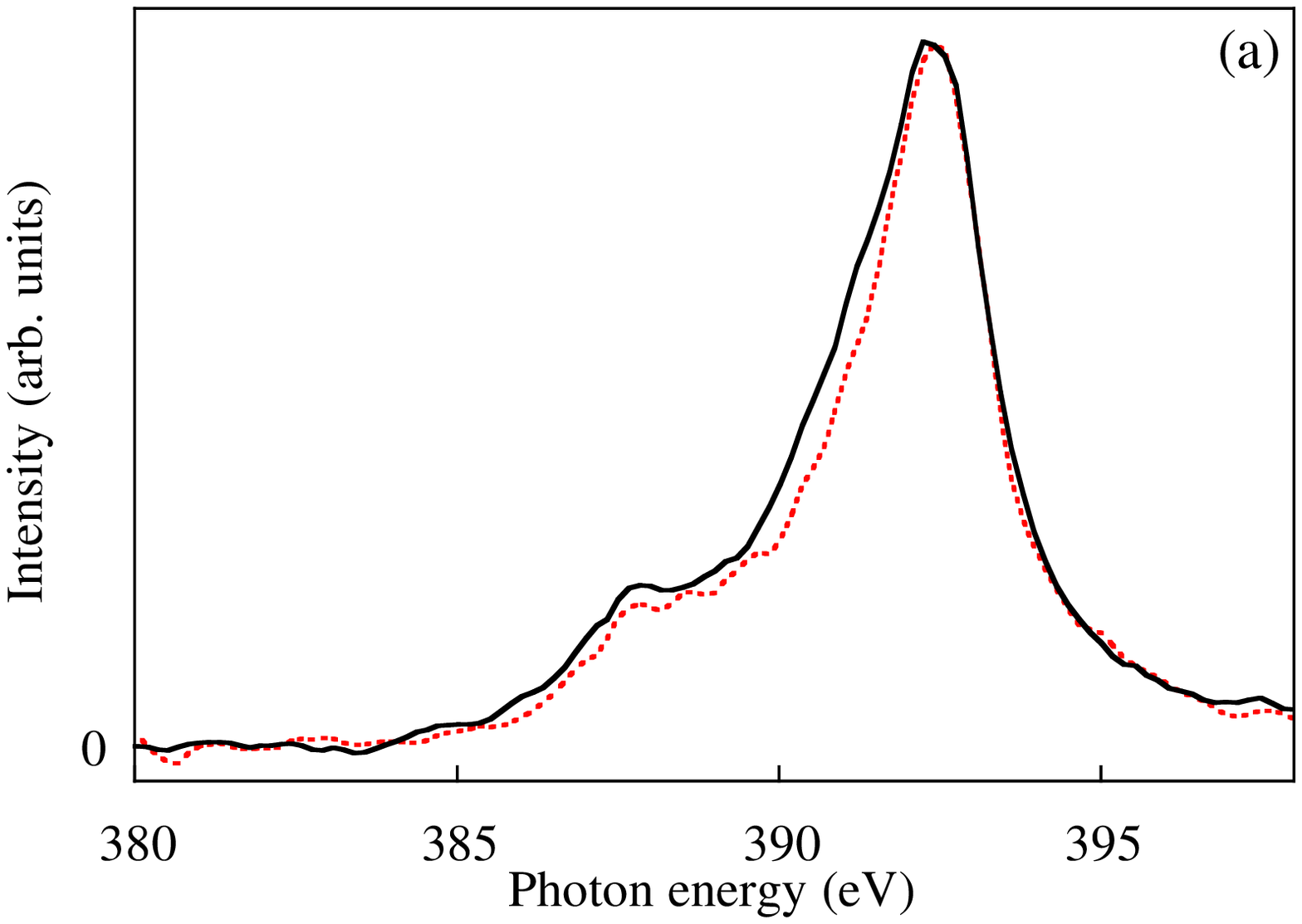}
\includegraphics[scale=0.75]{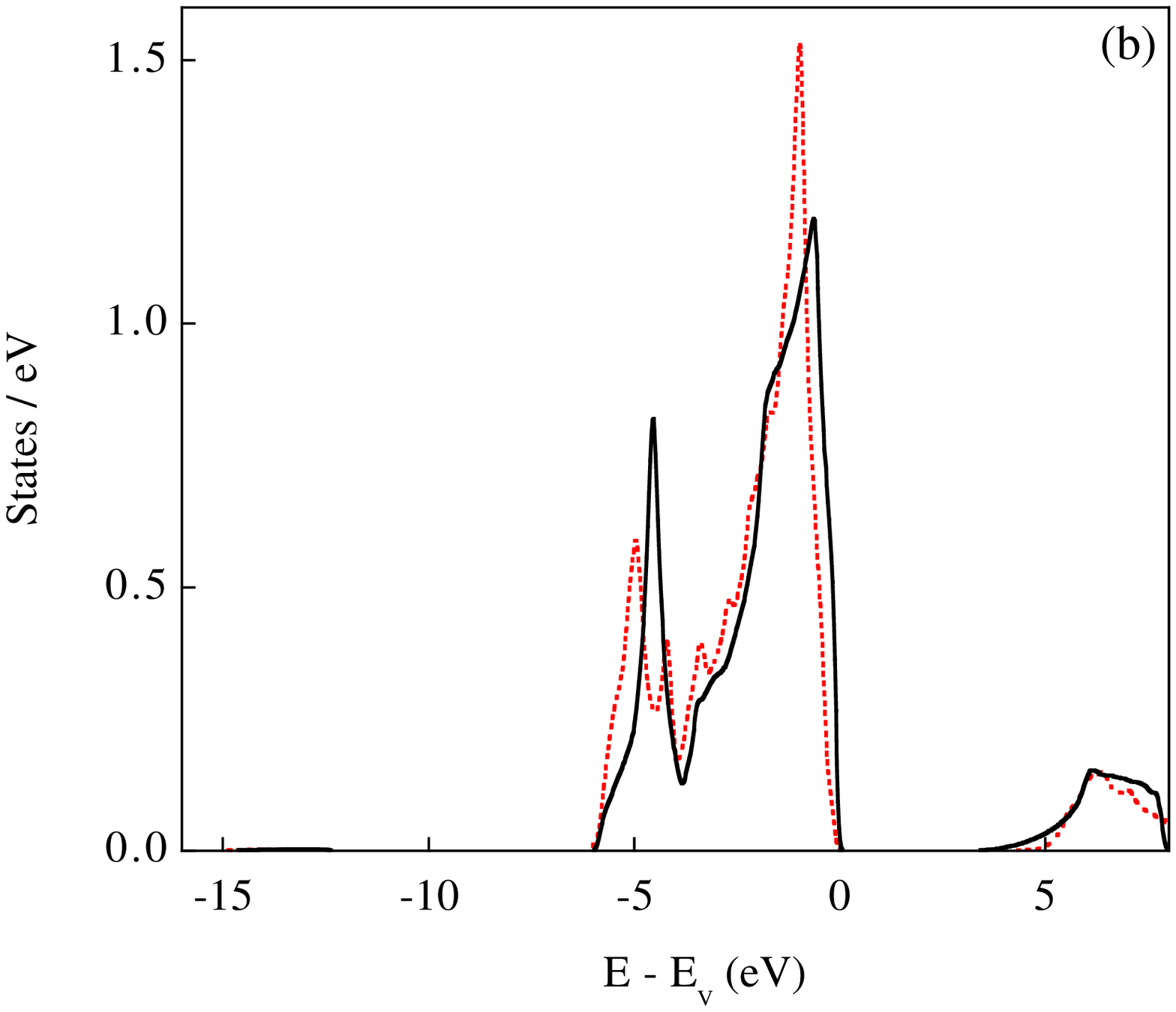}
\clearpage
\includegraphics[scale=0.75]{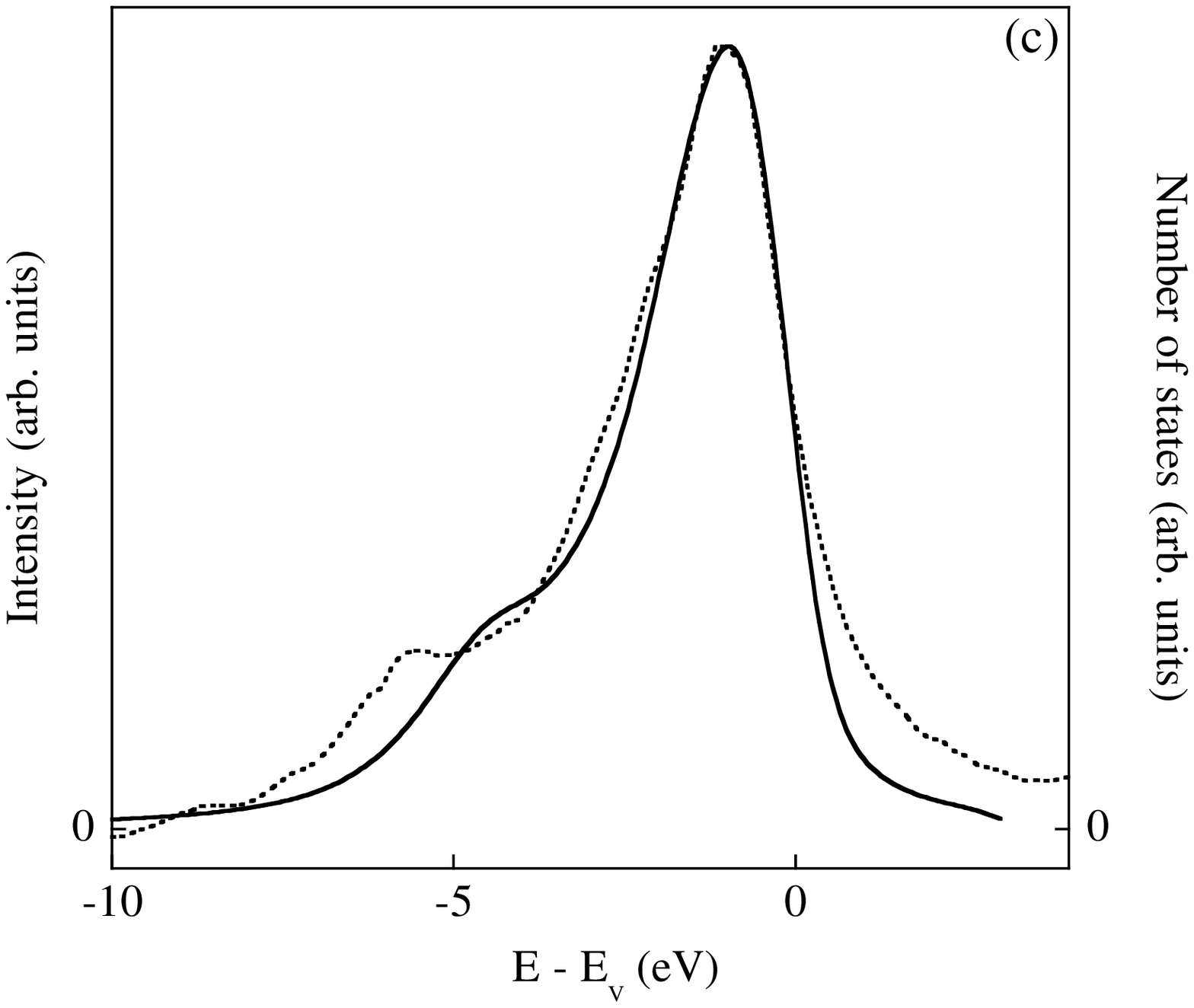}
\caption{(a) N $2p$ spectral distribution of AlN WZ (dotted line) and 
ZB (solid line) obtained with 3~keV incident electrons; (b) Theoretical 
N~$2p$ DOS of AlN WZ (dotted line) and ZB (solid line); (c) For AlN 
ZB, comparison of experimental (dotted line) and calculated (solid line) 
N~$2p$ spectra.}
\label{NK}
\end{figure}

The N~$2p$ spectral densities of WZ and ZB AlN are presented 
in Fig.~\ref{NK}a, on a photon energy scale. They are normalized with 
respect to their maximum. They present a peak at about 392~eV, having 
a shoulder around 388~eV. The spectrum of the wurtzite phase is in 
good agreement with the spectra found in the 
litterature~\cite{fom,smi,dud,klk}. With respect to the WZ phase,
the emission band is narrower towards the low photon energies
for the ZB phase. The valence band edges are the same for both phases.

The occupied theoretical N~$2p$ DOS presents a large maximum at a 
binding energy of about -1~eV and a second maximum at -5~eV. 
The difference between the WZ and ZB phases are small.
The calculations are in agreement with previous local and 
partial DOS calculated for AlN WZ and ZB~\cite{klk,lit}.

The comparison of the N~$2p$ spectral density with the calculated 
spectrum is shown in Fig.~\ref{NK}c for AlN~ZB. The region of the 
maximum of the experimental DOS is well reproduced by the calculation. 
However, the experimental shoulder energy is about 1~eV higher
than the calculated one. The DOS of AlN~WZ presents the same 
situation.

\section{Discussion}
The local atomic arrangement of the hexagonal WZ and cubic ZB AlN is as 
follows~\cite{mat}. In both phases the Al atoms are in a
tetrahedral environment. In the ZB phase, the distance between the Al 
atom and the first four N neighbors is 0.175~nm, and the $\widehat{AlNAl}$
angle formed by a N atom and two of its Al nearest neighbors is 
109.5$^\circ$ (perfect tetrahedron). In the WZ phase, there are two kinds 
of N nearest neighbors, surrounding the Al atoms inside a slightly 
distorted tetrahedron, located at 0.189 and 0.192~nm. This leads 
to two different $\widehat{AlNAl}$ angles of 117.7$^\circ$ and 107.7$^\circ$. 
There is also a difference concerning the second neighbor cation-anion 
bond distances~: 12 equivalents in ZB AlN, and 10 of three kinds in WZ AlN.

Since the XES is a local probe, the method is more sensitive to changes 
involving the nearest neighbors than to those due to differences 
in the long range interactions. When the structural environment
varies between two phases, pronounced changes of the DOS shapes are observed.
This is the case, for example, for $\alpha$-alumina (Al atoms in octahedral 
environment) and $\gamma$-alumina (Al atoms both in octahedral and tetrahedral 
environments)~\cite{kef}. For AlN, interactions between the first neighbors 
are about the same for the WZ and ZB phases, and one expects DOS changes 
to be smaller than those seen in the example cited above. In fact, no change 
is observed for the N~$2p$ spectral densities in the higher part of the 
valence band, and changes concerning the Al~$3p$ and Al~$3s$ spectral 
densities are limited. In fact, the changes mainly involve the Al states 
that are situated towards the lower binding energies, owing to their largely
extended character.

\begin{figure}
\includegraphics[scale=0.75]{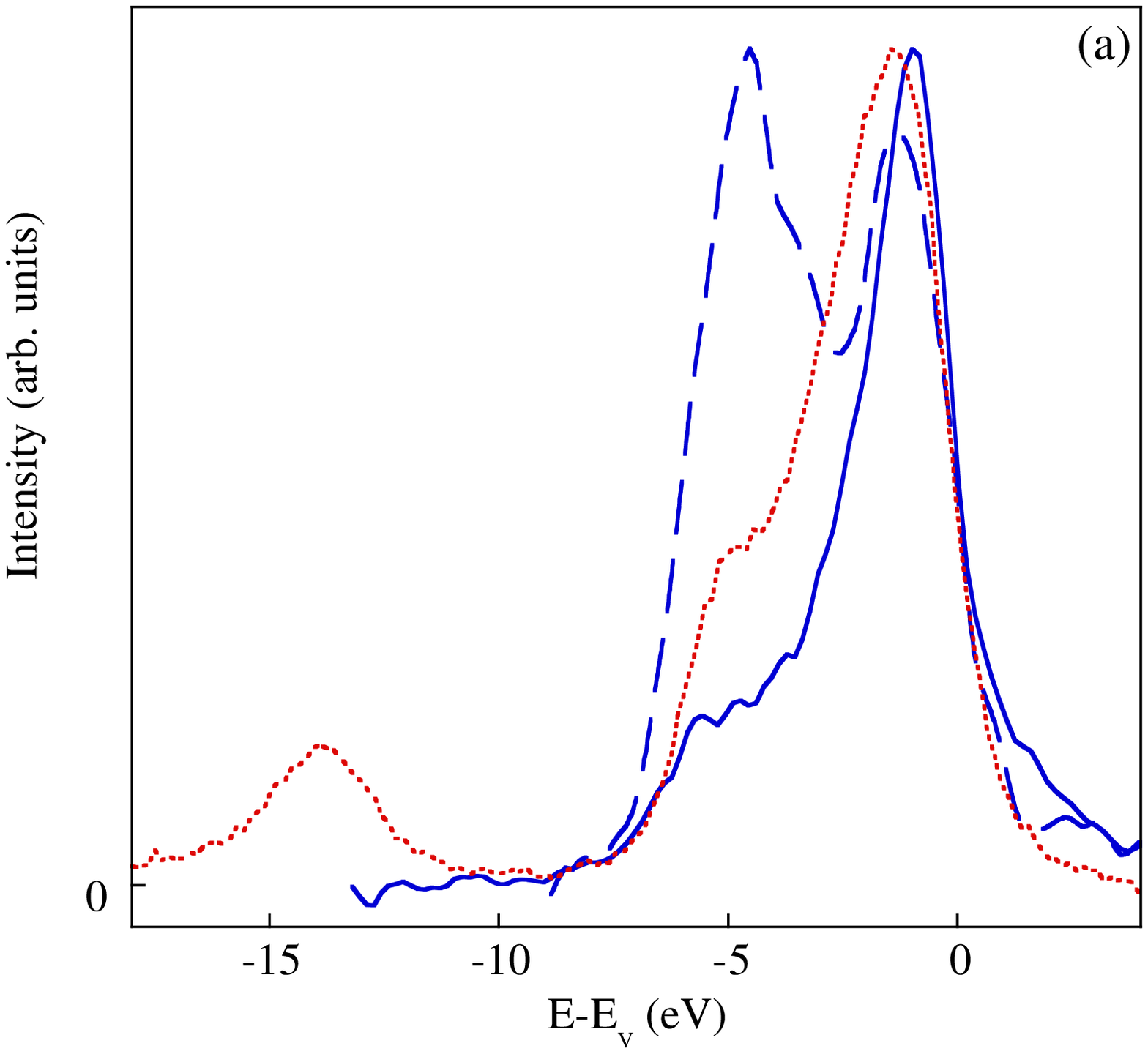}
\includegraphics[scale=0.75]{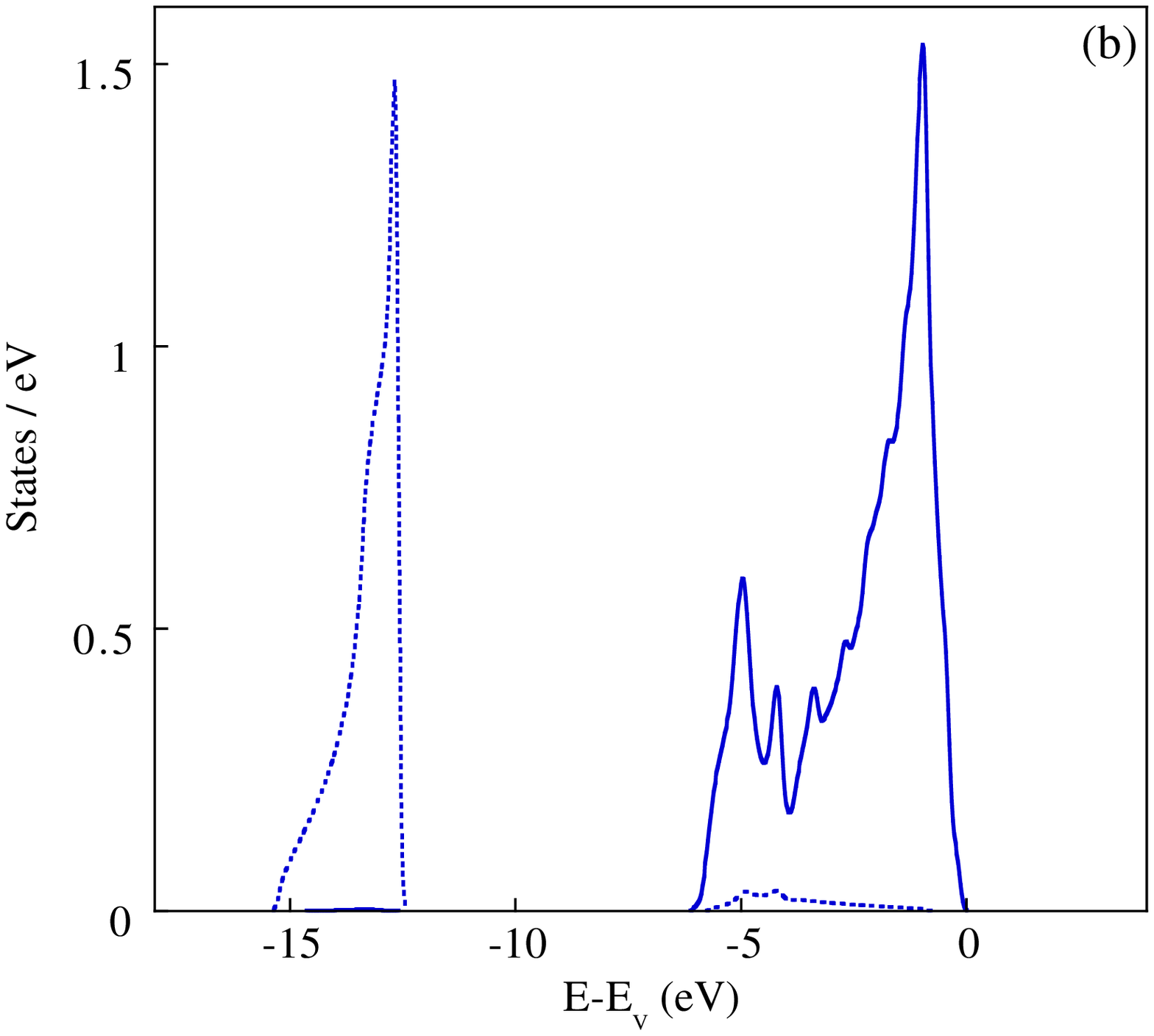}
\clearpage
\includegraphics[scale=0.75]{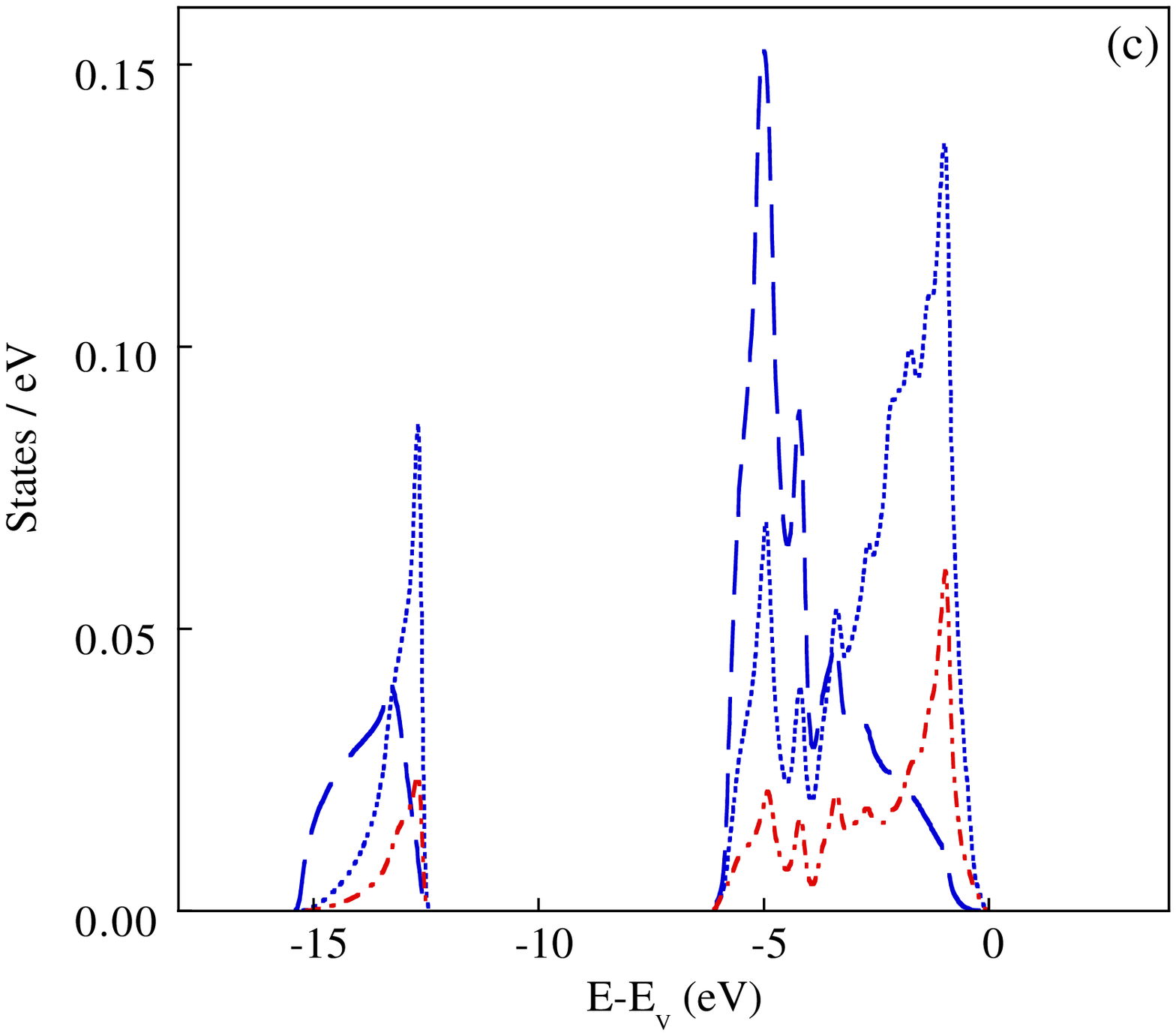}
\caption{For AlN WZ, comparison of the experimental and theoretical 
DOS. For sake of clarity the theoretical N and Al local DOS are presented 
separately. (a)~Experiment~: Al~$3sd$ states (dashed line); Al~$3p$ states (dotted line); 
N~$2p$ states (solid line). (b)~Theory~: N~$2s$ (dotted line); N~$2p$ 
(solid line). (c)~Theory~: Al~$3s$ states (dashed line); Al~$3p$ states 
(dotted line); Al~$3d$ states (dashed-dotted line).}
\label{BVexp}
\end{figure}

An important characteristic of the DOS for this kind of materials is 
the strong mixing between the various states constituting the valence band.
This is borne out in Fig.~\ref{BVexp}a, where the Al~$3p$, Al~$3sd$ and N~$2p$ 
spectral densities of WZ AlN are presented together on a binding energy 
scale relative to the top of the valence band. It can be seen that the 
occupied valence states are hybridized all along the valence band. 
Concerning the UVB, its top is dominated by the strongly 
mixed N~$2p$ and Al~$3p$ states, and its bottom by the 
N~$2p$ and Al~$3s$ ones. The maximum of the Al~$3s$ distribution and the 
shoulders of the Al~$3p$ and N~$2p$ distributions are in
correspondence around -6~eV. This energy distribution of the states 
within the valence band is in agreement with the DOS calculations, 
Fig.~\ref{BVexp}b and c.

In the region of the LVB, centered around -14~eV, only 
the Al~$3p$ states are observed. A maximum of the Al~$3s$ spectral density 
has been seen in Ref.~\onlinecite{wie} at the same experimental binding energy. 
According to Fig.~\ref{BVexp}b and c, all the different kind of states are 
hybridized in this region dominated by the N~$2s$ states.

Then, there exists a good agreement between the energies of the various 
features of the experimental and theoretical DOS. 
However, the K$\beta '$ structure is shifted with respect to the theory, 
Fig.~\ref{Alk}c. Such a shift has 
already been noted~\cite{ols,chi}, between experiments by x-ray photoelectron 
spectroscopy (giving the total DOS), and theoretical DOS. 
Another difference is observed between the 
N~K$\alpha$ emission and the N~$2p$ calculated spectrum, Fig.~\ref{NK}c. It has to 
be noted that this emission is quite weak, because it is observed at the 
third order of diffraction. This introduces a large uncertainty
in the feet of the emission that could account for the difference
with the theoretical DOS. However, the same kind of problem as the one occurring 
with the shift of the K$\beta '$ structure could explain this 
difference.

It is observed that the calculated spectra are narrower than 
the experimental ones (see Fig.~\ref{Alk}c, \ref{AlL}c and \ref{NK}c). It does 
not seem reasonable to use wider curves to broaden the theoretical DOS. 
Then, this discrepancy may be ascribed to an underestimation of the 
valence band width in the calculations. This would also explain the 
shift of the K$\beta '$ structure. Let us note, that x-ray and 
ultra-violet photoemission measurements of the valence band lead to an UVB width of 
8-9~eV~\cite{ols,gau,mar,wu}, whereas the theoretical width of the UVB is 
6~eV, in agreement with other calculations~\cite{chi,xu,lit}. So, 
even by taking into account a broadening of 1~eV to 
simulate the XPS experiments, it would not be possible to recover the 
correct experimental UVB width.

\section{Conclusion}
AlN thin films in the wurtzite and zinc-blende phases have been 
prepared by molecular beam epitaxy. The
electronic structures of AlN in the 
ZB and WZ phases have been determined experimentally by x-ray emission 
spectroscopy and calculated theoretically using the full potential linearized 
augmented plane wave method, within the framework of the density 
functional theory. It has been shown that the differences between both phases, 
in the experimental spectra or in the calculated DOS, are small because 
the two AlN structures are very similar. A relatively good agreement 
has been obtained between the observed and calculated energies of the various features close 
to the valence band maximum. However, some discrepancies appear at higher binding energies, 
which could be explained by an underestimation of the band widths in 
the calculations. Finally, it has been shown that the occupied valence 
states are hybridized all along the valence band.

\acknowledgments
The authors would like to thank Prof. J.-L. Schwob for his help 
concerning the x-ray grating spectrometer, and M. Krawczyk and the Centre 
de Calcul de Recherche et R\'eseau Jussieu for implementing 
the WIEN2k package. The authors are grateful with Pr. K. Schwarz for 
helpful remarks on the manuscript and Dr. P. Blaha for his help 
concerning the use of the WIEN2k package. Prof. C. Bonnelle is acknowledged 
for many fruitful discussions.

\begin{references}
\bibitem{jai} S. C. Jain, M. Willander, J. Narayan, and R. van 
Overstaeten, J. Appl. Phys. {\bf 87}, 965 (2000).
\bibitem{ver} F. Vergand, P. Jonnard, and C. Bonnelle, Europhys. Lett. 
{\bf 10}, 67 (1989); F. Vergand, P. Jonnard, M. Kefi, C. Bonnelle, C. Deparis,
and J. Massies, J. Phys. Condensed Matter {\bf 5}, 1691 (1993); P. Jonnard, 
F. Vergand, C. Bonnelle, M. Leroux, and J. Massies, Phys. Rev. B {\bf 55},
15727 (1997); and references therein.
\bibitem{hoh} P. Hohenberg and W. Kohn, Phys. Rev. {\bf 136},
B864 (1964); W. Kohn and L. J. Sham, Phys. Rev. {\bf 140}, A1133 (1965).
\bibitem{cha} T. Chassagne, G. Ferro, D. Chaussande, F. Cauwet, Y. Monteil, 
and J. Bouix, Thin Solid Films {\bf 402}, 83 (2002). 
\bibitem{mart} E. Martinez-Guerrero, E. Bellet-Almaric, L. Martinet, G. Feuillet, 
B. Daudin, H. Mariette, P. Holliger, C. Dubois, C. Bru-Chevallier, T. Chassagne, 
G. Ferro, and Y. Monteil, J. Appl. Phys. {\bf 91}, 4983 (2002).
\bibitem{bus} E. Bustarret, private communication.
\bibitem{bon1} C. Bonnelle, The Royal Society of Chemistry, Annual 
Report C, 1987, p. 201.
\bibitem{bon2} C. Bonnelle, F. Vergand, P. Jonnard, J.-M. Andr\'e, 
P.-F. Staub, P. Avila, P. Chargel\`egue, M.-F. Fontaine, D. Laporte, 
P. Paquier, A. Ringuenet, B. Rodriguez, Rev. Sci. Instrum. 
{\bf 65}, 3466 (1994).
\bibitem{jls} J.-L. Schwob, A. W. Wouters, S. Suckewer, and M. 
Finkenthal, Rev. Sci. Instrum. {\bf 58}, 1601 (1987).
\bibitem{w2k} P. Blaha, K. Schwarz, P. I. Sorantin, and S. B. 
Trickey, Comput. Phys. Commun. {\bf 59}, 399 (1990); K. Schwarz, 
P. Blaha, and G. K. H. Madsen, Comp. Phys. Commun. {\bf 147}, 71 
(2002); K. Schwarz and P. Blaha, Comput. Mat. Sci. {\bf 28}, 259 (2003).
\bibitem{lit} F. Litimein, B. Bouhafs, Z. Dridi, and P. Ruterana, N. J. 
Phys {\bf 4}, 64 (2002).
\bibitem{pbe} J. P. Perdew, K. Burke, and M. Ernzerhof, Phys. Rev. 
Lett. {\bf 77}, 3865 (1997).
\bibitem{ham} D. R. Hamann, Phys. Rev. Lett. {\bf 76}, 660 (1996).
\bibitem{sta} C. Stampfl and C. G. van de Walle, Phys. Rev. B {\bf 59}, 
5521 (1999).
\bibitem{pug} S. K. Pugh, D. J. Dugdale, S. Brand, and R. A. Abram, 
Semicond. Sci. Technol. {\bf 14}, 23 (1999).
\bibitem{mer} A. E. Merad, M. B. Kanoun, J. Cibert, H. Aourag, and G. 
Merad, Mat. Chem. Phys. {\bf 82}, 471 (2003).
\bibitem{fri} D. Fritsch, H. Schmidt, and M. Grundmann, Phys. Rev. B {\bf 
67}, 235205 (2003).
\bibitem{klk} K. Lawniczak-Jablonska, T. Suski, I. Gorczyca, N. E. 
Christensen, K. E. Attenkofer, R. C. C. Perera, E. M. Gullikson, J. H. 
Underwood, D. L. Ederer, and Z. Liliental Weber, Phys. Rev. B {\bf 61}, 
16623 (2000).
\bibitem{vur} I. Vurgaftman, J. R. Meyer, and L. R. Ram-Mohan, J. Appl. 
Phys. {\bf 89}, 518 (2001); I. Vurgaftman and J. R. Meyer, J. Appl. 
Phys. {\bf 94}, 3675 (2003).
\bibitem{wie} G. Wiech and E. Z\"{o}pf, J. Phys. {\bf 10}, C4-200 
(1971).
\bibitem{rei}J. Reisel, J. Geyer, J. Flock, and D. Sommer, X-Ray 
Spectrom. {\bf 32}, 29 (2003).
\bibitem{fom} V. A. Fomichev, Sov. Phys. - Sol. State {\bf 10}, 597 
(1968).
\bibitem{smi} K. E. Smith, L. C. Duda, C. B. Stagarescu, J. Downes, D. 
Korakakis, R. Singh, T. D. Moustakas, J. Guo, and J. Nordgren, J. Vac. 
Sci. Technol. B {\bf 16}, 2250 (1998).
\bibitem{dud} L. C. Duda, C. B. Stagarescu, J. Downes, K. E. Smith, 
D. Korakakis, T. D. Moustakas, J. Guo, and J. Nordgren, Phys. Rev. B
{\bf 58}, 1928 (1998).
\bibitem{mat} T. Mattila and A. Zunger, J. Appl. Phys. {\bf 85}, 160 (1999).
\bibitem{kef} M. Kefi, P. Jonnard, F. Vergand, C. Bonnelle, and E. 
Gillet, J. Phys. Condens. Matter {\bf 5}, 8629 (1993); S.-D. Mo and 
W. Y. Ching, Phys. Rev. B {\bf 57}, 15219 (1998).
\bibitem{ols} C. G. Olson, J. H. Sexton, D. W. Lynch, A. J. Bevelo, H. 
R. Shanks, B. N. Harmon, W. Y. Ching, and D. M. Wieliczka, Sol. State 
Commun. {\bf 56}, 35 (1985).
\bibitem{chi} W. Y. Ching and B. N. Harmon, Phys. Rev. 
B {\bf 34}, 5305 (1986).
\bibitem{gau} M. Gautier, J.-P. Duraud, and C. Legressus, J. Appl. Phys. 
{\bf 61}, 574 (1987).
\bibitem{mar} G. Martin, S. Strite, A. Botchkarev, A. Agarwal, A. 
Rockett, W. R. L. Lambrecht, B. Segall, and H. Morko\c c, J. Elec. Mat. {\bf 
24}, 225 (1995).
\bibitem{wu} C. I. Wu and A. Kahn, Appl. Phys. Lett. {\bf 74}, 546 (1999).
\bibitem{xu} Y.-N. Xu and W. Y. Ching, Phys. Rev. B {\bf 48}, 4335 (1993).
\end{references}
\end{document}